# 3D Topological Modeling and Multi-Agent Movement Simulation for Viral Infection Risk Analysis

Wassim Jabi[1], Yidan Xue[2], Thomas E. Woolley[3], Katerina Kaouri[4]

[1,2,3,4]Cardiff University

[1,2,3,4]{jabiw|xuey25|woolleyt1|kaourik}@cardiff.ac.uk

## Abstract

In this paper, a method to study how the design of indoor spaces and people's movement within them affect disease spread is proposed by integrating computer-aided modeling, multi-agent movement simulation, and airborne viral transmission modeling. Topologicpy spatial design and analysis software is used to model indoor environments, connect spaces, and construct a navigation graph. Pathways for agents, each with unique characteristics such as walking speed, infection status, and activities, are computed using this graph. Agents follow a schedule of events with specific locations and times. The software calculates "time-to-leave" based on walking speed and event start times, and agents are moved along the shortest path within the navigation graph, accurately considering obstacles, doorways, and walls. Precise distance calculations between agents are enabled by this setup. Viral aerosol concentration is then computed and visualized using a reaction-diffusion equation, and each agent's infection risk is determined with an extension of the Wells-Riley ansatz. Infection risk simulations are improved by this spatio-temporal and topological approach, incorporating realistic human behavior and spatial dynamics. The resulting software is designed as a rapid decision-support tool for policymakers, facility managers, stakeholders, architects, and engineers to mitigate disease spread in existing buildings and inform the design of new ones. The software's effectiveness is demonstrated through a comparative analysis of cellular and open commercial office plan layouts.

**Keywords**: Computer-Aided Modelling, Spatial Design and Analysis, Topological Analysis, Navigation Graphs, Multi-Agent Movement Simulation, Aerosol Distribution, Viral Transmission, Infection Risk.

# 1. Introduction

The COVID-19 pandemic has prompted researchers to study how the configuration of indoor spaces and the activities taking place within them affect the risk of viral transmission [1–5]. Lessons learned from the pandemic have prompted new architectural design guidelines for infection prevention and control [6].

Historically, major disasters have often led to significant changes in building regulations. For instance, the Great Fire of London in 1666, which destroyed a large portion of the city, prompted the implementation of stricter fire safety standards, including the use of brick and stone in construction instead of wood, and the creation of wider streets to prevent the rapid spread of future fires [7]. Another example is the 1906 San Francisco earthquake and subsequent fire, which led to improvements to the water supply infrastructure and the development of more stringent building codes to improve structural integrity and earthquake resistance [8].

Similar to earlier outbreaks such as the spread of measles [3], the COVID-19 outbreak has revealed the link between building attributes and infection risk [1]. This highlights the urgent need to analyse building designs and develop decision-support tools for policymakers, stakeholders, architects, and engineers to help them design and implement changes that mitigate infection risk. More specifically, there is a need for simulation models that balance the requirements for accuracy and speed of execution, take into consideration complex geometries, can simulate people movement according to daily schedules and event locations, and visualize scenarios and the resulting infection risk in a manner easily accessible to stakeholders.

To address these needs, this research aims to create an accessible web-based decision-support tool for optimizing building designs to enhance safety and health in response to current and future pandemics. To achieve this aim, we propose to integrate computer-aided 3D modelling of spaces, navigational graphs, agent-based simulation, and mathematical models of the viral concentration and infection risk. We hypothesize that this integrated approach using a realistic representation of indoor spaces, multi-agent movement simulation, and spatio-temporal mathematical models provides a more accurate analysis of infection risk and more readily support informed decision-making in building design.

## 2. Related Work

Various methodologies and theoretical frameworks inform our analysis of indoor space design and people movement in the context of viral transmission. By examining these diverse yet interconnected fields, we aim to provide a robust foundation for understanding and mitigating the risk associated with viral transmission in indoor environments. Related areas of research include non-manifold topological modelling [9], dual graphs [10], space syntax [11], multi-agent simulation [12], and viral transmission modelling [3,13,14]. Although viral transmission modelling plays a significant role in formulating the theoretical underpinnings and quantitative predictions of this research, a detailed exposition is beyond the primary scope of this paper and is provided in a companion paper under preparation [15].

### Non-manifold Topological Representations

Most 3D modelling software relies on an underlying modelling 'engine' that provides the basic functionality and needed computations to compose and edit 3D geometry. These engines are usually classified as either manifold or non-manifold [16]. Manifold Topology (MT), or more precisely 2-manifold topology, refers to bodies without internal voids that can be fabricated from a single block of material. A key property of manifold objects is that a sphere drawn centred on any point of the object's surface will be divided into two distinct regions: one inside and one outside the object. This clear division is a fundamental characteristic that makes manifold bodies manufacturable and physically realizable.

However, manifold modelling lacks the direct ability to derive adjacency information or the presence of shared entities as each manifold object exists independently of other objects. While commercial 3D modelling software systems that rely on manifold engines are readily available for sophisticated geometric modelling, they fall short as extendable simulation engines as they do not usually model topological connections within indoor spaces, and do not simulate people movement, making it challenging to meet the outlined needs.

In contrast, Non-Manifold Topology (NMT) is defined mathematically as cell complexes within Euclidean space, representing spatial relationships between geometric entities [17]. Practically, NMT allows any combination of vertices, edges, surfaces, and volumes to exist in a single logical body, where multiple cells can share faces, faces can meet at an edge or multiple edges at a vertex. Coincident vertices, edges, and faces are merged.

While NMT results in configurations that cannot be physically realizable, this characteristic provides the direct ability to find adjacencies between regions or spaces, enabling the construction of 'dual' navigational graphs that can be easily traversed and searched for shortest paths. A dual graph is a graph that represents the adjacency relationships of a geometric or topological structure. Each region (e.g. a room or volume) of the original graph is represented as a node in the dual graph. An edge exists between two nodes in the dual graph if their corresponding regions in the original graph share a common boundary. These links can have various weights or attributes assigned to them. This network can be employed for indoor navigation, where paths between two points can be determined using graph algorithms. For example, non-manifold topology and dual graphs have been used to build an emergency response tool that accurately computes egress paths in case of an emergency [18].

## Graph Theory and Space Syntax

Airborne viral transmission in indoor environments requires the simulation of how people within that environment move and interact with each other. A person's physical proximity and duration of interaction with another infected person directly affects their risk of contracting an airborne virus. Whitehead and Eldars, for example, empirically mapped the movements of nurses within a theatre suite for one day's duty creating a 'string' diagram showing the routes taken and the heaviest concentrations of movements [19]. Their approach aggregates the movements over time to arrive at the heaviest trafficked areas. In contrast, our approach calculates viral concentration at each simulation time step taking into consideration the location of agents at that time step.

The need to understand and map how people use and navigate the built environment has developed into the well-established field of space syntax [20]. The original definition of space syntax constitutes objects, relations, and operations with two kinds of entities: a solid entity (obstacles), and a vacant entity through which movement is possible. Given this framework, space syntax evolved to modelling and analysing the built environment, road networks, and whole cities [21]. It has also moved beyond 2D representations to analysing 3D configurations [22]. Through graph-theoretic approaches, space syntax measures have shown correlations with human spatial behaviour and have, thus, been used to predict the likely impacts of architectural and urban spaces on users.

## Multi-Agent Simulation

Multi-agent simulations model the behaviour and interactions of individuals moving within a shared space. This approach helps in predicting movement patterns and contact points that could enhance or mitigate airborne viral transmission. One of the foundational works in agent-based systems is Carl Hewitt's "Actor Model" introduced in 1973 [12]. The Actor Model conceptualized independent entities (actors) that interact with each other through message passing. This model laid the groundwork for concurrent computation and influenced the development of agent-based systems.

In 1996, Epstein and Axtell introduced the concept of agent-based computer programming to the study of social phenomena such as transmission of culture and disease propagation [23]. Within the Built Environment research area, Turner and Penn studied pedestrian behaviour using agent-based systems. They found that by imbuing many agents with attributes and parameters such as destination, field of view, and location between events, they can generate movement patterns like those found in the real world [24]. Schuamann et al. also used agent-based system for pre-occupancy evaluation of architectural designs [25]. Their model can analyse how different spatial configurations impact human spatial behaviour.

During the COVID-19 pandemic, agent-based models with stationary agents and no architectural structure (a rectangular domain) were used to compare non-pharmaceutical interventions (NPIs). For example, Moore et al. and Woodhouse et al (2022) used agent-based models to compare the infection risk in educational indoor settings [26,27].

Ozcan and Haciomeroglu implemented a path-based multi-agent navigation model that uses the A* shortest-path algorithm to handle local trajectories while global decisions are handled by a global planner [28]. We follow a similar approach but use Dijkstra's shortest path algorithm instead [29].

Balkan et al. simulated infection risk in indoor spaces using schedules and multi-agent movement [30]. Their navigation model is based on NOMAD, a microscopic pedestrian movement model. The NOMAD model provides a detailed and realistic simulation of pedestrian behaviour by focusing on individual decision-making, strategic interactions, and the optimization of movement within a discretized environment. While the NOMAD model offers a highly detailed and realistic approach to simulating pedestrian movements, its computational intensity, and complexity of implementation, can limit its practicality in certain applications.

Hernandez-Mejia et al also used a multi-agent approach to study how architectural interventions can mitigate the spread of SARS-CoV-2 in emergency departments [31]. Their model, based on records from a German hospital, uses a time manager to organize patient arrivals and manage the duration of their stay in each area of the emergency department. One of the limitations of their approach is that their model of the emergency department is based on a flowchart of activities and events rather than a realistic architectural floor plan with accurate locations for agents. Our choice to use a simpler navigation model within an accurate architectural floor plan balances accuracy and computational efficiency which is a key challenge to implementing fast simulations as those necessary in a fast-evolving epidemic.

Ciunkiewicz et al. developed a simulation framework to forecast COVID-19 transmission in localized environments [32]. They followed a similar approach to ours using agents, scenarios, and a universal scheduler to move agents from one position to another and calculate, at each time step, their infection risk. However, their 2D spatial map is more basic than ours in that it uses a chessboard-like matrix where agents can move only orthogonally or diagonally. In contrast, our spatial maps are geometrically realistic with shortest navigation paths that do not follow a pre-determined grid and can turn corners and avoid obstacles in a realistic manner. Given that we use a fully three-dimensional spatial model, we can also determine if there are barriers between two agents. This is important as not to mistakenly assume, for example, that two agents are close to each other when, in fact, they are on either side of a wall.

Finally, while sophisticated commercial people movement simulation engines, such as those used for simulating crowd egress in case of fires are readily available (Gwynne & Kuligowski, 2009), they are usually domain-specific desktop software packages that are not easily integrated in an open and bespoke web-based system. Thus, we found them not suitable to use for our purposes.

## Mathematical Modelling of Airborne Viral Transmission

Mathematical models of airborne diseases have been used to estimate infection risk since the pioneering work by [14] and [3]. In the Wells-Riley model, the infectious aerosols are assumed to be well-mixed in the indoor space, neglecting spatial heterogeneity due to architectural details and the behaviour and interactions of individuals.

Here, we simulate the concentration of the infectious aerosols emitted by infectious individuals using a reaction-diffusion equation. The infection risk of individuals is determined using a spatio-temporal infection risk formula [4] which extends the Wells-Riley ansatz by incorporating also spatial information. Our new model determines the number of moving individuals being infected and those remaining susceptible to the disease as time progresses, assuming the "dose" of viral particles needed for a person to get infected. An explanation of the mathematical modelling details is beyond the scope of this paper and will be included in a companion paper which provides useful insights and guidance to policymakers and space managers through comparing and ranking a series of NPIs according to the type of indoor space [15].

## 3. Methodology

The proposed software system consists of several critical modules. Firstly, a geometric and topological editor is necessary for representing the geometry of indoor spaces and their connectivity, enabling both topological and spatial analysis of buildings and their internal spaces. Using this editor, a navigational graph is constructed that supports the placement and simulation of multiple agents moving through the building. An agent-specific schedule, with a series of events at varying locations in a building, provides the needed information to accurately locate the agents at each time step. Lastly, calculating spatio-temporal infection risk is made feasible through an extension of the Wells-Riley ansatz.

The integration of these modules allows us to simulate different scenarios of practical interest using diverse building typologies and mixtures of agents inhabiting and moving through them. Further, the software offers a comprehensive and accessible decision-support tool for optimizing building designs to enhance safety and health in response to current and future public health challenges.

## Spatial Modelling and Analysis Software

We leveraged topologicpy, an open-source and extendable topological 3D spatial modelling and analysis environment [33,34]. Through its reliance on a non-manifold modelling engine for boundary representation, topologicpy can precisely model the required geometry, automatically build the topological connections among spaces, and answer spatial and topological queries. The topologicpy application protocol interface (API) contains nine (9) core classes (Vertex, Edge, Wire, Face, Shell, Cell, CellComplex, Cluster, and Topology), seven (7) secondary classes (e.g. Aperture, Dictionary, Graph, Grid, Helper, Matrix, Vector), and several additional utility and specialized classes (Figure 1). Of note here is the Dictionary class that can store keys and values and be associated with any topology. Dictionaries allow us to store and retrieve attributes as and when needed. topologicpy's API is extensive, covering areas of geometric construction, spatial analysis, visualization, energy simulation, and artificial intelligence among other capabilities [35]. In this paper we focus on the capabilities and classes that facilitate our work.

The most basic topology in topologicpy is a Vertex which represents a point in 3D space with X, Y, and Z coordinates. Two vertices can be connected by an Edge. Several Edges can be connected in various ways to form a Wire (a polygon or a non-manifold network of edges). A polygonal, planar, closed Wire can form a Face which has a direction vector and an area. Faces can have holes in them (called internal boundaries). Several Faces that share edges can be combined into a Shell (an equivalent to a mesh) which can be either open or closed. A manifold shell that is closed forms a Cell which has a volume. Several Cells that share faces can be combined into a CellComplex. A Cluster (a group) is an abstract class that has any number of other topologies in it. Finally, a Topology is an abstract super-class of all other types of topologies that contains methods that can apply to multiple types of topologies.

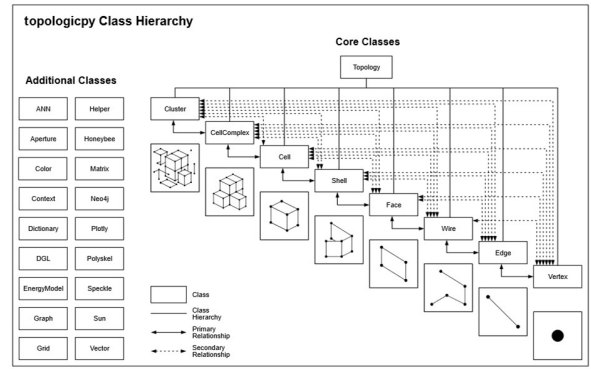

*Figure 1. The topologicpy class hierarchy.*

We use topologicpy's ability to create 'primitives' such as lines, rectangles, circles, and prismatic solids to create our architectural models (Figure 2). topologicpy can also import 3D models created using other 3D modelling software through standard file formats.

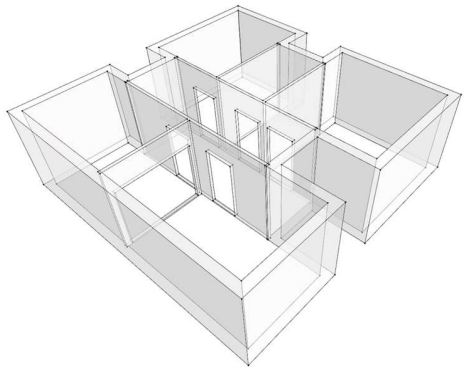

*Figure 2 3D model created in topologicpy.*

In addition, topologicpy provides several types of spatial and topological queries:

1. Constituent Query: This type of query allows the user to retrieve the constituent sub-topologies that make up a topology. For example, the user can ask topologicpy to provide the faces, the edges, and the vertices of a cell.

2. Hierarchical Query: This type of query can be thought of as the reverse of a constituent query. This query allows the user to retrieve the super-topologies that have the queried topology as part of their constituent sub-topologies. For example, the user can ask an edge for the faces that it is a part of. Or they can ask a face for a list of cells that it is a part of.

3. Lateral Query: This type of query is used for finding adjacent topologies. For example, a user can ask a cell to return the list of cells that share a face with that cell.

4. Containment Query: This type of query allows the user to retrieve the contents of a topology. Unlike constituent parts that make up the geometry of a topology, contents are independent topologies that are simply stored in a topology (think furniture in a room as opposed to the walls and corners of a room).

5. Intersectional Query: This type of query allows the use to find what two topologies share. For example, a user can ask two neighbouring cells for the list of faces, edges, or vertices that they share.

6. Associative Query: This type of query allows the user to find topologies linked to another topology through dictionary information rather than topological proximity. Each topology in topologicpy has a dictionary with varying number of keys and values stored in it. These dictionaries can contain references to other topologies (e.g. by an ID key). This is akin to building a relational database or a graph that can be traversed to retrieve the needed information.

We used these methods to decompose and analyse the 3D model and derive the needed information to create a simplified 3D model with furniture obstacles represented as simple prisms (Figure 3).

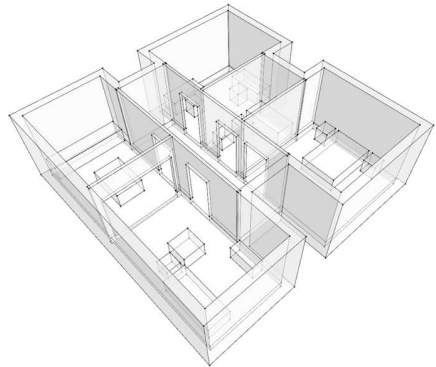

*Figure 3 3D model of an architectural indoor environment with furniture.*

## Navigation Graph

The next step in our methodology is to create the navigation graph that controls the paths on which agents travel. A navigation graph is a representation of a spatial environment where nodes correspond to locations (vertices) and edges represent possible paths or connections between these locations [36]. This graph structure is used to model the navigability of a space, allowing for efficient route planning within that environment. The API of topologicpy includes the ability to directly create a navigation path from a face with an optional list of obstacles that are converted into holes in the face. Thus, in the context of a building floor plan, walls, furniture and other obstacles are subtracted from the floor plate to create a face with holes (Figure 4). As can be seen on the left side of Figure 4, We slightly lowered the furniture vertically to subtract it from the floor plate using topologicpy's Boolean operations. This created the needed holes in the face, as seen on the right side of Figure 4. If a buffer area between the navigable space and obstacles is required, the user can use topologicpy's Face and Wire offsetting methods to create larger holes and to offset the exterior boundary inwards (Figure 5).

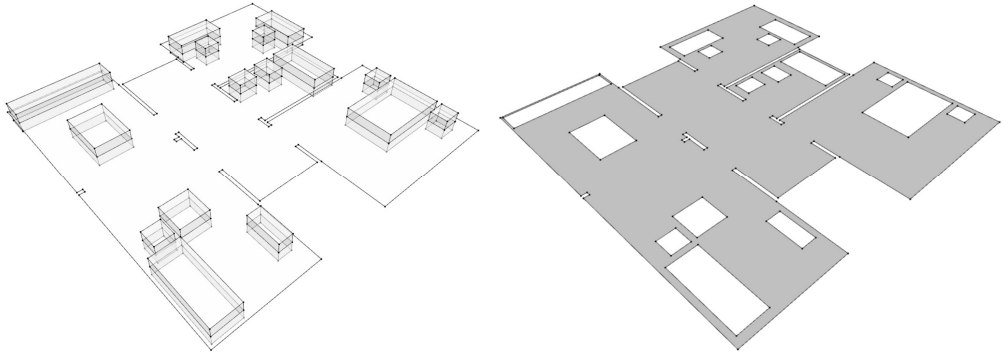

*Figure 4 Derving the host face with holes from a floor plan with obstacles.*

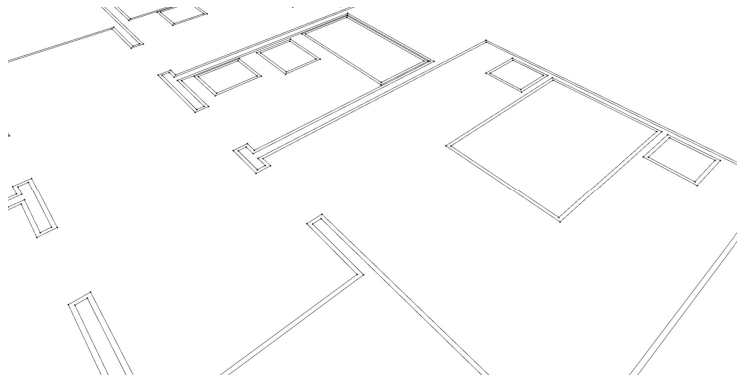

*Figure 5 Offsetting the boundaries to create a buffer area between the navigable space and the obstacles.*

The navigation graph algorithm uses parallel processing to connect a list of source vertices to a list of destination vertices on a face without traversing over holes in that face. The algorithm first checks if the distance between the source and destination vertices exceeds a specified tolerance. If the distance is less than or equal to the tolerance, the connection is discarded. This ensures that only connections between vertices that are significantly apart are considered. Then, an edge is created between the source and destination vertices. A Boolean intersection operation is performed between this edge and the face (the floor plan of the building with obstacles represented as holes in the face). The edge is only considered valid if the result of this Boolean operation is also an instance of an Edge class rather than a null result or a cluster of edges. A null result means the edge is either completely outside the face or completely within a hole in the face. A cluster of edges result means that the edge has been split into disjointed edges by one or more holes in the face. Filtering for results that are an instance of the Edge class ensures that the edge intersects fully with the face and is not outside its boundaries or intersecting in an invalid manner (Figure 6). The final list of valid edges is sent to the Graph class to create the navigation graph. The resulting graph can be visualised by asking it for its topology (wire) and displaying that topology (Figure 7).

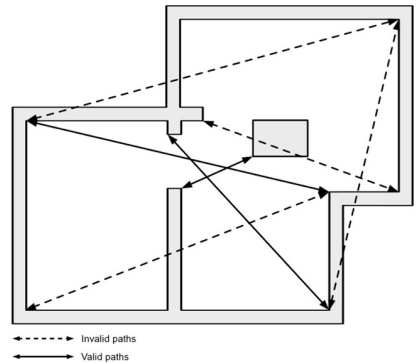

*Figure 6 Example of valid and invalid paths.*

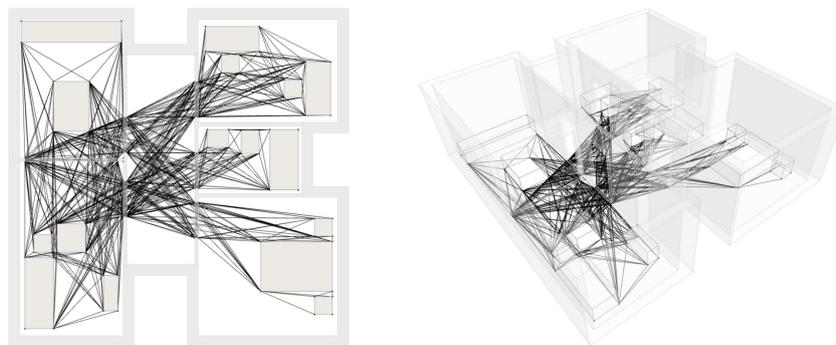

*Figure 7 The created navigation graph. Plan view (left), Perspective view (right).*

For larger floor plans, the creation of the navigation graph might become prohibitively time-consuming. In that case, an alternative graph made of a pruned Delaunay triangulation might be sufficient to represent how agents move in an indoor environment (Figure 8). This process is detailed below in the illustrative case study.

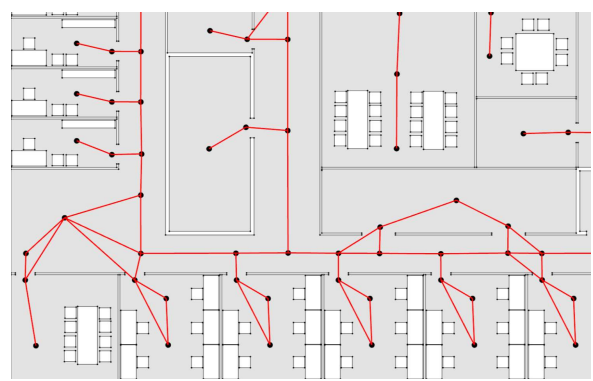

*Figure 8 Example of a simplified navigation graph in an office plan using a pruned Delaunay triangulation.*

## Events, Activities, Schedules, Agents

Our multi-agent simulation system models events, schedules, and agents. An event is defined by a location in the building where an activity takes place for a certain duration of time. It is represented using (X, Y, Z) coordinates. A schedule is a list of events each associated with its own start time and the type of activity taking place. In this research, an agent's level of physical and vocal activity can vary. This affects their breathing rate and aerosol emission rate, and, thus, has a significant impact on the infection risk value.

The scheduling system does not allow time gaps. Thus, the duration of an event is calculated based on the subtraction of the start time of the next event from its own start time.

A daily schedule is then assigned to an agent which are represented by a user-specified geometry (e.g. a cylinder or a cone) with a local (X, Y, Z) origin. Each agent has several attributes that are stored in its dictionary that are necessary for the simulation calculations such as:

- Walking speed: determines how fast an agent moves.
- Infection status: indicates if the agent is infectious, infected but not yet infectious, or susceptible.
- Superspreader: defines whether the agent is a superspreader or not (only applies to infectious agents).
- Mask: states whether if, and what type of, masks are worn, e.g. cotton, surgical or N95.
- Activity type: states where the agent is resting, talking, talking loudly, moderately exercising, or vigorously exercising.

Agents navigate a spatial environment following a pre-calculated navigation graph. To travel from one event to another, the system uses topologicpy's Graph class to compute the shortest path between any two points on it, using Dijkstra's shortest path algorithm (Dijkstra, 1959). A path is an instance of the Wire class which can compute the position of a vertex on it given a user-specified distance from its starting point. At each time step, the agent's new position is computed as:

$$V = Wire.VertexByDistance(P, T \times S) \qquad (1)$$

where

$V$ is the calculated vertex position.
$P$ is the calculated shortest path.

$T$ is the current time.
$S$ is the agent's walking speed.

An agent remains at its last location until the current time step is equal to or greater than a computed time-to-leave parameter. This parameter is computed as follows:

$$T = A - \left(\frac{L}{S}\right) \qquad (2)$$

where:

$T$ is the time-to-leave parameter.
$A$ is the desired arrival time at the next event.
$L$ is the path length.
$S$ is the agent's walking speed.

## Simulation and Visualization

The viral aerosol concentration is determined using a reaction-diffusion equation, which is implemented via a fast finite element method [37]. The infection risk of each individual, as a function of time, is estimated using an extension of the Wells-Riley ansatz [4]. The average infection risk is calculated by the sum of the infection risk of each susceptible individual, divided by the number of susceptible individuals.

The simulation results are visualised in topologicpy as a heatmap projected on the floor faces of the 3D model using the Plotly graphing library which is integrated into topologicpy [38]. At each timestep, the heatmap, the location and colours of the agents are updated in the 3D model as well as the simulation results. These frames can then be automatically saved and compiled into an animation.

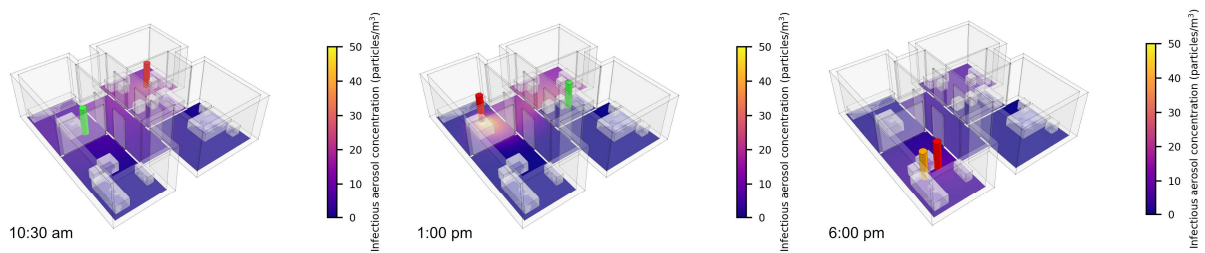

*Figure 9 Simulation results showing the aerosol concentration in a household (heatmap) and agent locations at three times during the day. The infectious agent is represented by a red cylinder. The susceptible agent is represented by a green cylinder when not infected, and an orange one when infected.*

# 4. Illustrative Case Study

To illustrate the features of the proposed system, we studied two hypothetical office plan layouts. The first layout emphasizes cellular offices, whereas the second layout favours an open-plan design (Figure 10). Both layouts fit into the same rectangular floor plate with an overall width of 40 m and overall length of 20 m. The layouts have the same dual core where the lifts and staircases are located. It is important to note that the layouts presented here are fictitious. Any resemblance to actual office plans is purely coincidental.

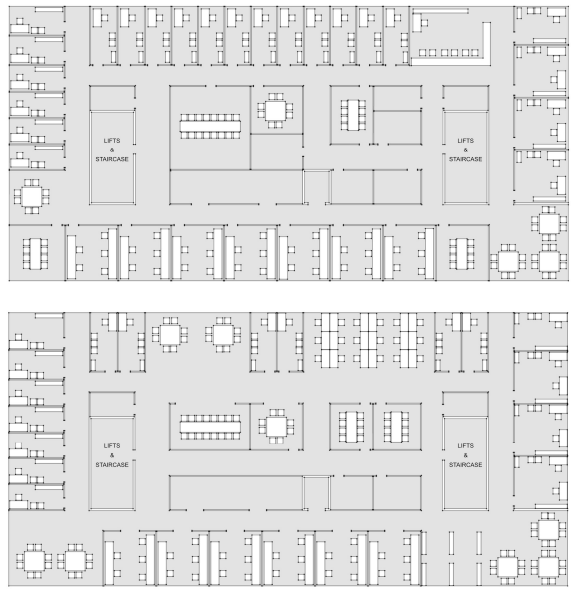

*Figure 10 Two hypothetical office plan layouts. Top: cellular layout, Bottom: open-plan layout.*

## Constructing the Navigation Graph

The next step was to construct the navigation graph for each layout. Given the size and complexity of the floor plan, constructing a standard navigation graph, as described above, is computationally expensive. Thus, an alternative semi-automated approach is followed. The layout is populated with a set of points representing significant locations as follows: Firstly, the centre of each chair in the layout was added to the list of points. Secondly, two points were added to either side of each door opening. Thirdly, a set of boundary points were added around or near each piece of furniture to allow access. Fourthly, a point was placed near the centre of each cellular office and other defined areas. Lastly, corridors were populated with colinear points to allow navigation along their main axis. These can be junctures at which a person would decide to turn in a different direction (Figure 11).

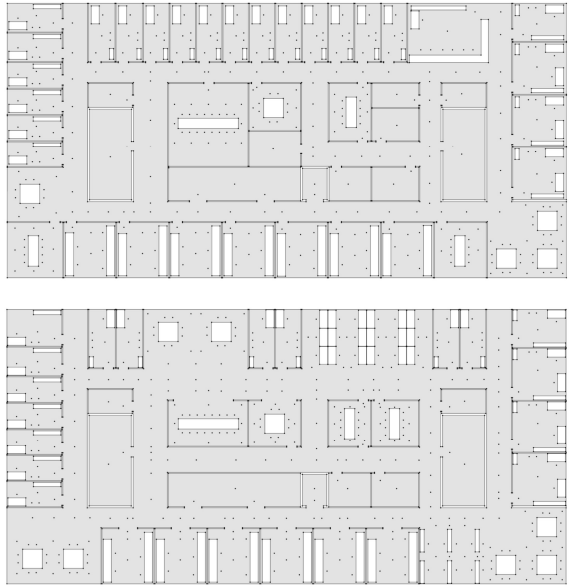

*Figure 11 Significant point locations for both layouts. Top: cellular layout, Bottom: open-plan layout.*

The points were then triangulated using a Delaunay algorithm [39] that divides their convex hull into triangles, where the circumcircles of these triangles do not enclose any of the points (Figure 12).

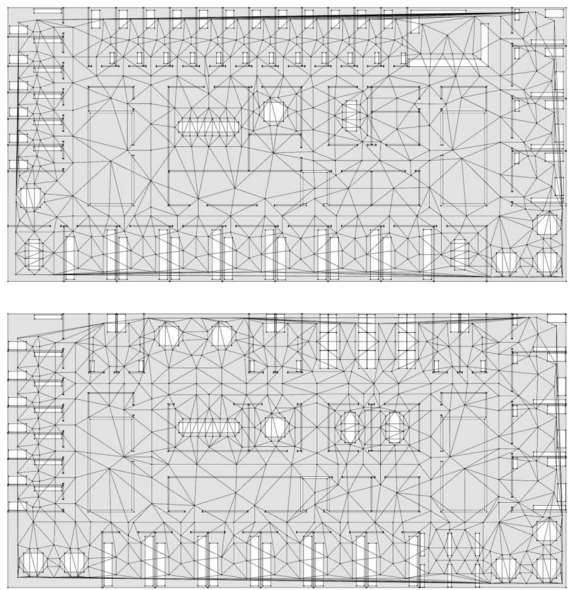

*Figure 12 Delaunay triangulation for both layouts. Top: cellular layout, Bottom: open-plan layout.*

To derive the navigation graph, each edge in the Delaunay triangulation was intersected with the walls and obstacles in the floor plan. Only edges that do not intersect any walls or obstacles are kept (Figure 13). These filtered edges and their end vertices are then used as input to create a navigation graph using the *Graph.ByVerticesEdges()* method. All information is then transferred to the multi-agent simulation software scripts.

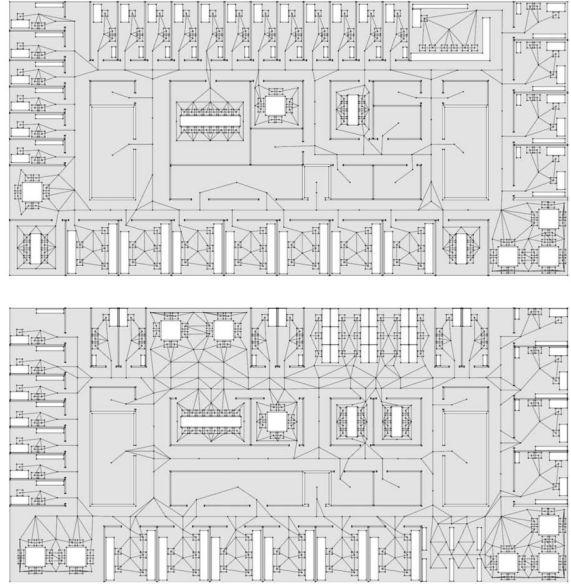

*Figure 13 The final navigation graph for each layout. Top: cellular layout, Bottom: open-plan layout.*

## Multi-agent Simulation

We populated the two multi-agent simulation scenarios (cellular plan, open plan) with 60 agents each. Agents are assumed to be familiar with the office layout and thus always travel along shortest paths within the navigation graph using their assigned walking speed. We gave both scenarios the same overall schedule as follows:

1. 9:00 am: All agents arrive from two lift cores and walk to their assigned desk location. Starting at 9:30 am, 20 agents visit 1 to 6 other offices, staying at each location for 5-10 minutes.
2. 11:00 am-11:30 am: 40 agents meet in conference rooms.
3. 1:00 pm-2:00 pm: Agents have lunch in three shared spaces. A first group has lunch from 1:00 pm to 1:30 pm and a second group has lunch from 1:30 pm to 2:00 pm.
4. 2:00 pm-3:00 pm: All agents work at their assigned desk.
5. 3:00 pm-3:30 pm: 40 agents meet in conference rooms.
6. 3:30 pm-5:00 pm: All agents return to their assigned desk location. 20 agents are allowed to visit 1 to 4 other offices from 3:30 pm, staying at each location for 5-15 minutes.
7. 5:00 pm: All agents depart through the two lift cores.

The simulation settings were also kept the same for both scenarios as follows:

1. Ventilation: very poor, with 0.12 air changes per hour (ACH).
2. Environment: constant temperature, relative humidity and sunlight, which result in a exponential decay rate of the viruses in infectious aerosols.
3. Walking speed: 1.5 metres per second.
4. Infection rate: 10% of the agents (i.e. 6 agents) are infectious, while 90% of these are susceptible. No infectious agent is a superspreader.
5. Agent activities: Each agent is walking when moving, talking loudly during meetings, talking when having lunch or visiting other offices, and not talking when working at their office desk. The aerosol emission rate of an infectious agent decreases in the order listed above for these four (4) activities (walking, talking loudly, talking, and resting).

## Results

The highest concentrations of infectious aerosols were observed in meeting rooms and lunch areas, which aligns with the expected outcome due to the extended periods of proximity and interaction of the individuals in these spaces (Figure 14).

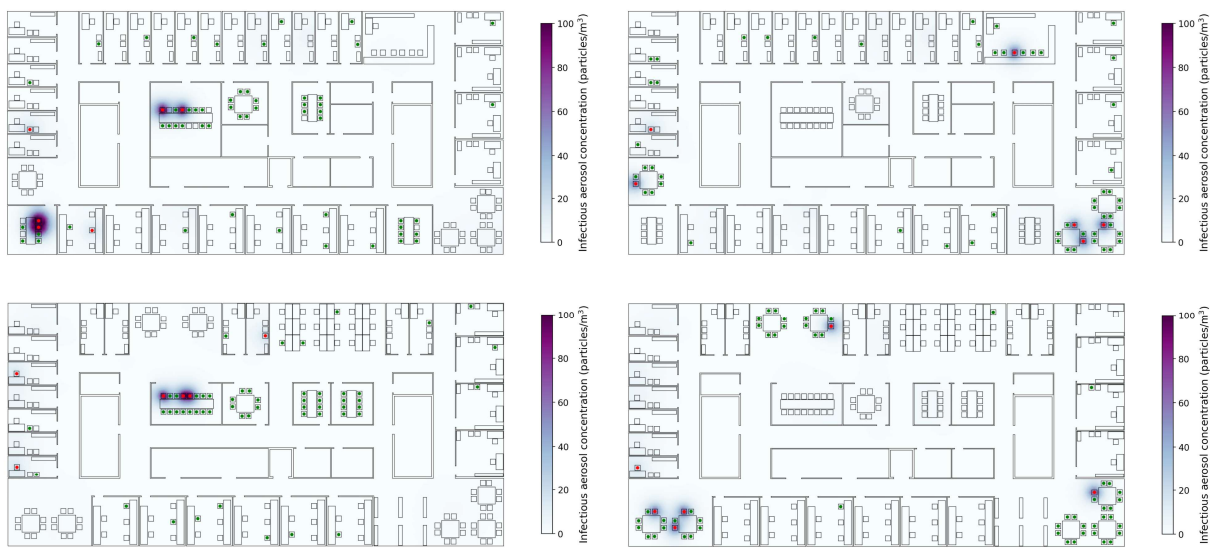

*Figure 14 Areas of highest infectious aerosol concentration for each layout. Top: cellular layout, Bottom: open layout. Left: Meeting rooms, Right: Lunch areas,*

The results indicate, in this case, and with these inputs, that between the hours of 9:00 am and 3:30 pm, the cellular office layout is slightly more effective in reducing the infection risk compared to the open-plan layout. After 3:30 pm, however, the infection risk in the cellular office layout becomes slightly higher than in the open-plan layout and continues to be so until 5:00 pm (Figure 15). The presence of dividing walls and fewer opportunities for communal interaction in the cellular layout helps isolate agents by minimizing their exposure to infectious aerosols. However, it is important to note that the purpose of this singular illustrative study is not to provide any quantitative evidence or interventions. Rather, its aim is to illustrate how this methodology can be used effectively to compare different spatial configurations and to inform the design decision-making process for architects and space managers.

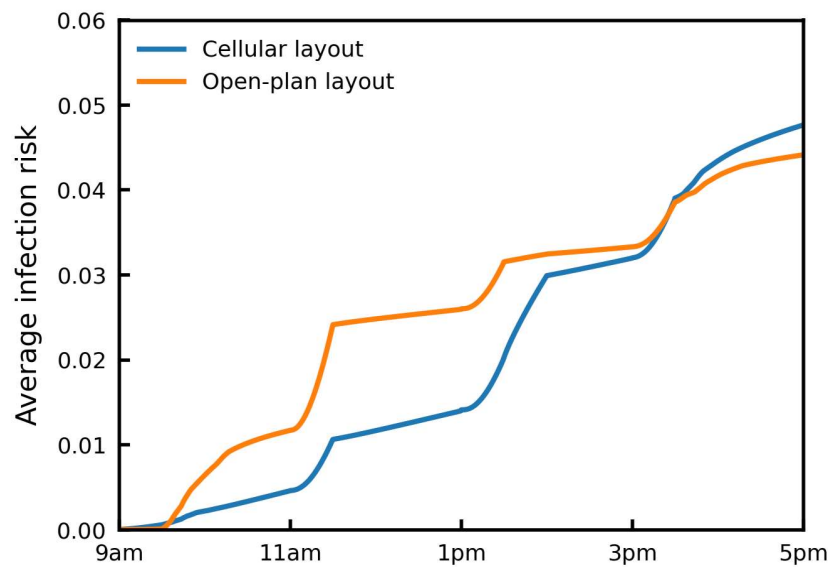

*Figure 15 Average infection risk of susceptible individuals for the two layouts.*

## 5. Limitations and Future Work

The primary limitation of this study is that the simulation does not use real-world data to model variations in human movement but instead relies on idealised linear movements using shortest paths. While the supporting software offers a fully 3D environment, the modelling of aerosol concentrations is in two dimensions for speed considerations. Additionally, the model does not account for all possible environmental factors, such as natural and mechanical ventilation systems. Thus, future research should aim to incorporate more comprehensive environmental variables, such as HVAC systems, and to validate simulations with real-world data. Expanding the scope to include different types of buildings and spaces, as well as varying demographic and behavioural patterns, would enhance the robustness and applicability of this study.

We are in the final stages of completing a web app that guides users on how to upload 3D models, add events, schedules, and agents, specify various parameters, and visualize the results. The impact of this app will not only affect the field of architectural design but also public health policy and facility management. The resulting web app will also be accessible to the public and managers of indoor settings, allowing them to design safer spaces and reduce disease spread. It also offers policymakers data-driven insights for creating guidelines that promote healthier spaces.

## 6. Concluding Remarks

This study provides a novel, multidisciplinary approach that integrates detailed 3D modelling of spaces, agent-based navigation modelling, and spatio-temporal modelling of the viral aerosol concentration and infection risk for susceptible individuals. The office layout case study clearly illustrates that spatial configuration can significantly influence infection risk in indoor environments. As we continue to confront complex global health threats, such innovative multidisciplinary research methodologies will be essential in developing effective strategies to prevent future epidemics.

## Disclosure statement

## Data availability statement

The data that support the findings of this study are available from the corresponding author, WJ, upon reasonable request.

## Funding

This work was supported by the UK Engineering and Physical Sciences Research Council (EPSRC) grant EP/X525522/1 for Cardiff University.